\documentclass{jpp}
\usepackage{graphicx}
\usepackage{epstopdf, epsfig}
\usepackage{eurosym}
\usepackage{graphicx}
\usepackage{caption,subcaption}
\usepackage{float}
\usepackage[normalem]{ulem}
\usepackage{color}
\usepackage{soul}

\usepackage{bbold}
\usepackage{amsmath,amssymb}
\usepackage{braket}
\usepackage[english]{babel}
\usepackage{ulem}
\usepackage{cancel}

\newcommand{\beq}{\begin{equation}}
\newcommand{\eeq}{\end{equation}}

\newcommand{\bu}{{\bf u}}

\newcommand{\bk}{{\bf k}}
\newcommand{\bq}{{\bf q}}

\newcommand{\bx}{{\bf x}}

\newcommand{\beqa}{\begin{eqnarray}}
\newcommand{\eeqa}{\end{eqnarray}}

\newcommand{\bb}{{\bf b}}

\def\be{\begin{equation}}
\def\ee{\end{equation}}

\shorttitle{Magnetic reconnection, plasmoids and numerical resolution}
\shortauthor{J.M.G. Morillo \& A. Alexakis}

\title{ Magnetic reconnection, plasmoids and numerical resolution }

\author{José María García Morillo \& Alexandros Alexakis \aff{1}
\corresp{\email{alexakis@phys.ens.fr}}
   }

\affiliation{Laboratoire de Physique de l’Ecole Normale Supérieure, ENS, Université PSL, CNRS, Sorbonne Université, Université Paris-Diderot, Sorbonne Paris Cité, Paris, France }

\begin{document}

\maketitle

\begin{abstract}

Explaining fast magnetic reconnection in electrically conducting plasmas has been a theoretical challenge 
in plasma physics since its first description by Eugene N. Parker. In the recent years the observed reconnection rate 
has been shown by numerical simulations to be explained by the plasmoid instability that appears in highly conductive plasmas. 
In this work we show that the plasmoid instability is very sensitive to the numerical resolution used. 
It is shown that well resolved runs display no plasmoid instability even at Lundquist number as large as $5\cdot10^5$
achieved at resolutions of $32\,768^2$ grid points.
On the contrary in simulations that are under-resolved below a threshold, the plasmoid instability manifests itself 
with the formation of larger plasmoids the larger the under-resolving is. 
The present results thus question the description of the plasmoid instability as a mechanism for fast magnetic reconnection.

\end{abstract}

\section{Introduction    }    
\label{sec:intro}            

Magnetic reconnection refers to the sudden change of magnetic topology due to Ohmic dissipation or other micro-scale plasma processes. 
In astrophysics it is met in solar flares, coronal mass ejections, the solar wind and the Earth’s magnetosphere to mention a few examples.
In laboratory scales it is observed in tokamak discharges, and in reversed field pinch devices. 
It is responsible for the fast acceleration of charged particles and plasma heating \cite{yamada2010magnetic}.
It was noted early on \cite{giovanelli1946theory} that the rate of reconnection observed in astrophysical plasmas was much faster
than the relevant Ohmic time scale. The model of Sweet and Parker \citep{parker1957sweet,sweet1958electromagnetic} improved on this estimate 
by introducing what is now known as the Sweet-Parker model where the reconnection timescale is accelerated by a factor
of $\sqrt{S_L}$ where $S_L$ stands for the Lundquist number defined as $S_L=V_A L/\eta $ where $V_A$ is the Alfven speed, $L$ 
the typical structure size and $\eta$ the magnetic diffusivity. Although the Lundquist number in astrophysical plasmas
is large, the improvement of the Sweet-Parker model still lacks orders of magnitude compared to observations. 
Different, explanations have been put forward to produce faster reconnection rates than the Sweet-Parker model
including different geometry of the layer \citep{petschek196450}, Hall effect \citep{morales2005hall,wang2000collisionless},  
Electron pressure \citep{egedal2013review,wang2000collisionless}, Electron inertia \citep{andres2014effects} and turbulence \citep{lazarian2015turbulent,lazarian20203d}. 
However even without adding additional physics it has been argued that a two dimensional magnetohydrodynamic (2D-MHD)
model as the one proposed by Sweet-Parker can result in {\it fast} ({\it ie} magnetic diffusivity independent) reconnection rate 
if the Lundquist number is large enough so that the plasmoid instability develops \citep{shibata2001plasmoid}.
The plasmoid instability appears
for $S_L \gtrsim 10^4$
\citep{loureiro2007instability,samtaney2009formation} and leads to the formation of magnetic islands along the current sheet that enhance the reconnection rate \citep{lapenta2008self,bhattacharjee2009fast,samtaney2009formation,
daughton2009transition,cassak2009scaling,huang2010scaling,
huang2012distribution,loureiro2013fast,huang2013plasmoid,
loureiro2012magnetic,uzdensky2010fast}. 
These results were based on extended numerical simulations using a variety of codes including particle in cell methods, finite volume and pseudospectral methods.
We argue however in this work that reconnection is particularly sensitive to the numerical resolution and some of these results would need to be reexamined.

\section{Numerical model}   

\begin{figure}                                                                                   
  \centerline{\includegraphics[width=0.48\textwidth]{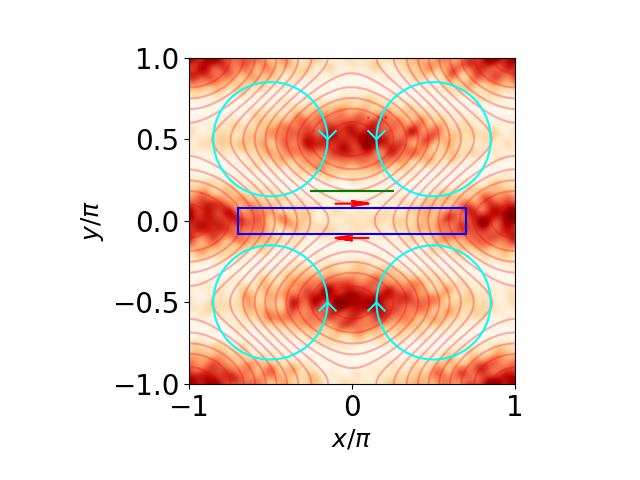} 
           \includegraphics[width=0.48\textwidth]{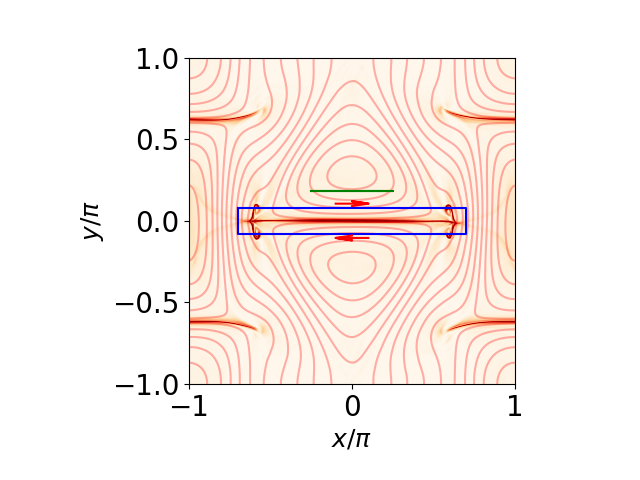} }
  \caption{Visualisation of the current density of the initial conditions and the resulting current layer in the entire domain.
  Red lines, indicate the magnetic field lines while blue lines indicate the velocity field. The blue box marks the zoomed in region 
  that is shown in the subsequent figures.
  }
\label{fig:IC}
\end{figure}

In this work we revisit the reconnection problem in 2D-MHD paying particular emphasis on numerical convergence. 
We consider the 2D-MHD equations in a double periodic square box of size $L=2\pi$.
In terms of vorticity and the magnetic vector potential they read:
\beqa 
\partial_t  \omega + \bu\cdot \nabla \omega &=& \bb \cdot \nabla j + \nu \nabla^2 \omega \\
\partial_t  a + \bu\cdot \nabla a &=&  \eta \nabla^2 a
\eeqa 
where $\omega = {\bf e}_z \cdot \nabla\times \bu$ is the vorticity with ${\bf e}_z$ the direction perpendicular to the examined plane and ${\bu}$ the velocity field. The magnetic field is given by $\bb=\nabla\times ({\bf e}_z a)$ where ${\bf e}_z a$ 
is the magnetic vector potential. The current along ${\bf e}_z$ is given by $j={\bf e}_z\cdot \nabla \times \bb = -\nabla^2 a$. The viscosity $\nu$ is set equal to the magnetic diffusivity $\eta$ for all our simulations. 
%
As initial conditions we consider the Orsang-Tang vortex \citep{orszag1979small} plus a small perturbation: 
\beq 
a(t=0,\bx )    =  A_0 [-\cos(x)+\cos(2y)/2 ]  + a_p
\eeq
while the velocity field is defined by its stream function $\psi$ 
(such that $u_x=\partial_y \Psi$ and $u_y=-\partial_x \Psi$) by  
\beq 
\Psi(t=0,\bx ) =  \Psi_0 \sin(x)\sin(y) + \psi_p. 
\eeq 
The amplitudes $A_0$ and $\Psi_0$ are such that the initial magnetic energy
density is $\frac{1}{2}\langle |\bb|^2 \rangle=\frac{1}{2}$  and the kinetic energy is 
$\frac{1}{2}\langle |\bu|^2 \rangle=\frac{1}{8}$.
The perturbations $a_p,\psi_p$ are chosen to include Fourier modes with 
wavenumber $|\bk|<16$ with random phases and their amplitude are such that
their energy corresponds to $0.25\%$ of the total energy. They provide a seed for
linear instabilities to develop that otherwise would depend on the the round-off error.
A visualisation of the initial conditions in terms of the current square is shown in 
the left panel of figure \ref{fig:IC}. 

The equations were solved using the {\sc ghost} pseudospectral code \citep{mininni2011hybrid} 
with a 4th order Runge-Kutta scheme for the time advancement, 
the 2/3 rule for de-aliasing and
using a uniform grid of $N$ grid points in each direction. 
Many different numerical simulations were carried out varying the resolution and the value of $\eta=\nu$. 
The parameters of all our runs are given in the table \ref{tab:table2}.

\begin{table}
\begin{center}
\begin{tabular}{||ccc||ccc||ccc||}
 $N$   &       $\eta$  &           $S_L$ & $N$    &      $\eta$  &           $S_L$ & $N$    &      $\eta$  &           $S_L$ \\
\hline
{\bf 1024 } &    0.50E-01  &     0.19E+02 & {\bf  2048} &     0.20E-02  &     0.12E+04 & {     4096} &     0.15E-04  &     0.14E+06 \\ 
{\bf 1024 } &    0.20E-01  &     0.70E+02 & {\bf  2048} &     0.10E-02  &     0.26E+04 & {     4096} &     0.70E-05  &     0.35E+06 \\ 
{\bf 1024 } &    0.10E-01  &     0.18E+03 & {\bf  2048} &     0.50E-03  &     0.54E+04 & {     4096} &     0.50E-05  &     0.46E+06 \\ 
{\bf 1024 } &    0.50E-02  &     0.41E+03 & {\bf  2048} &     0.20E-03  &     0.15E+05 & {     4096} &     0.20E-05  &     0.91E+06 \\ 
{\bf 1024 } &    0.20E-02  &     0.12E+04 & {\bf  2048} &     0.10E-03  &     0.30E+05 & {     4096} &     0.10E-05  &     0.18E+07 \\ 
{\bf 1024 } &    0.10E-02  &     0.26E+04 & {     2048} &     0.50E-04  &     0.61E+05 & {\bf  8192} &     0.50E-04  &     0.62E+05 \\ 
{\bf 1024 } &    0.50E-03  &     0.54E+04 & {     2048} &     0.20E-04  &     0.13E+06 & {\bf  8192} &     0.15E-04  &     0.18E+06 \\ 
{\bf 1024 } &    0.20E-03  &     0.15E+05 & {     2048} &     0.15E-04  &     0.17E+06 & {     8192} &     0.10E-04  &     0.28E+06 \\ 
{    1024 } &    0.10E-03  &     0.30E+05 & {     2048} &     0.50E-05  &     0.24E+06 & {     8192} &     0.50E-05  &     0.51E+06 \\ 
{    1024 } &    0.50E-04  &     0.47E+05 & {     2048} &     0.30E-05  &     0.39E+06 & {     8192} &     0.25E-05  &     0.95E+06 \\ 
{    1024 } &    0.30E-04  &     0.59E+05 & {     2048} &     0.20E-05  &     0.58E+06 & {     8192} &     0.20E-05  &     0.12E+07 \\ 
{    1024 } &    0.20E-04  &     0.87E+05 & {     2048} &     0.15E-05  &     0.75E+06 & {     8192} &     0.10E-05  &     0.23E+07 \\ 
{    1024 } &    0.15E-04  &     0.10E+06 & {     2048} &     0.10E-05  &     0.14E+07 & {\bf  16384} &     0.10E-04  &     0.28E+06 \\ 
{    1024 } &    0.10E-04  &     0.14E+06 & {\bf  4096} &     0.20E-03  &     0.15E+05 & {    16384} &     0.50E-05  &     0.39E+06 \\ 
{    1024 } &    0.70E-05  &     0.16E+06 & {\bf  4096} &     0.10E-03  &     0.30E+05 & {    16384} &     0.25E-05  &     0.86E+06 \\ 
{    1024 } &    0.50E-05  &     0.23E+06 & {\bf  4096} &     0.50E-04  &     0.62E+05 & {\bf 32768} &     0.50E-05  &     0.54E+06 \\ 
{    1024 } &    0.15E-05  &     0.61E+06 & {     4096} &     0.30E-04  &     0.11E+06 &             &               &              \\ 
\hline
\end{tabular}
\caption{ Simulation parameters $N,\eta,S_L$. Boldface $N$ is used for well-resolved and marginally well resolved runs. }
\label{tab:table2}
\end{center}
\end{table}


The evolution of the system leads to the formation of a current sheet aligned along the $x$-axis
centered at $x=0$. The intensity of the current sheet measured by the mean current density squared
$\langle j^2 \rangle $ increases rapidly and peaks at a time around $t\simeq 1.9$ after which it decays. 
In what follows all the studies are performed at the peak of $\langle j^2 \rangle $. 
At this time we define the Lundquist number as $S_{L}\equiv B_{max}/(\eta k_1)$
where $k_1=1$ is the smallest non-zero wavenumber. To calculate $B_{max}$ 
for each $y$ we calculate the mean magnetic field $\overline{b_x}(y)$ along the $x$ direction 
in the range $x\in [-\pi/8,\pi/8]$ (shown by the horizontal green line in figure \ref{fig:IC}).
$B_{max}$ is then defined as the first local maximum of $\overline{b_x}(y)$ as one moves away
from the current sheet at $y=0$. The non-dimensional reconnection rate is defined here 
as $RR=u_{in}/B_{max}$ where $u_{in}$ is again calculated by finding the mean 
inwards velocity $-\overline{u_y}(y)$ over the same segment as for $\overline{b_x}(y)$
and then $u_{in}$ is defined as the first maximum of $-\overline{u_y}(y)$ as one moves away from the current layer. 
Note that this average is crucial in the presence of plasmoids that
make local vales of $u_y$ and $b_x$ fluctuate strongly.

\section{Results}                                                

Exact solutions of reconnection layers describing the formation of the reconnection 
are not feasible and one needs to rely on numerical solutions.
For the validity of a numerical method to be verified one needs to demonstrate that for a given set of physical parameters
there exists a resolution $N_c$  such that all larger resolutions $N>N_c$ give the same result, up to a small error that can be bounded by a decreasing function of $N$. 
Such a procedure proves that the the numerical solution does not depend on the resolution and approaches the exact solution of the problem. 
Different numerical methods have different convergence rates. 
Finite difference and finite volume codes lead to  a power-law convergence 
implying that the error made decreases as a negative power-law as $N>N_c$ is increased, 
while pseudo-spectral and finite element codes result in an exponential convergence. 
This exponential convergence can be realised by considering the energy spectrum of the involved fields 
here defined as $E_b(k)=\frac{1}{2}\sum_{k<|\bq|\le k+1} |\tilde{\bf b}_\bq|^2 $
where $\tilde{\bf b}_\bq$ is the Fourier transform of the  magnetic field $\bf b$. Similarly, the squared current spectrum is defined as $E_J(k)=k^2E_b(k)$.
For a smooth field the energy and current spectrum display an exponential decrease with the wavenumber at large $k$.
Further increase of resolution thus adds exponentially small corrections.
In the present study we have considered that a simulation is well resolved if the peak of the squared current spectrum defined as $E_J(k)=k^2E_b(k)$ is at least ten times larger than its value at $k=k_{max}=N/3$ the maximum allowed wavenumber ie $\max_k\{ E_J(k)  \} \ge 10  E_J(k_{max})$. This implies that most of the Ohmic dissipation is correctly captured. The consequences of violating this criterion are severe.

\begin{figure}                                                                                   
  \centerline{\includegraphics[width=0.9\textwidth]{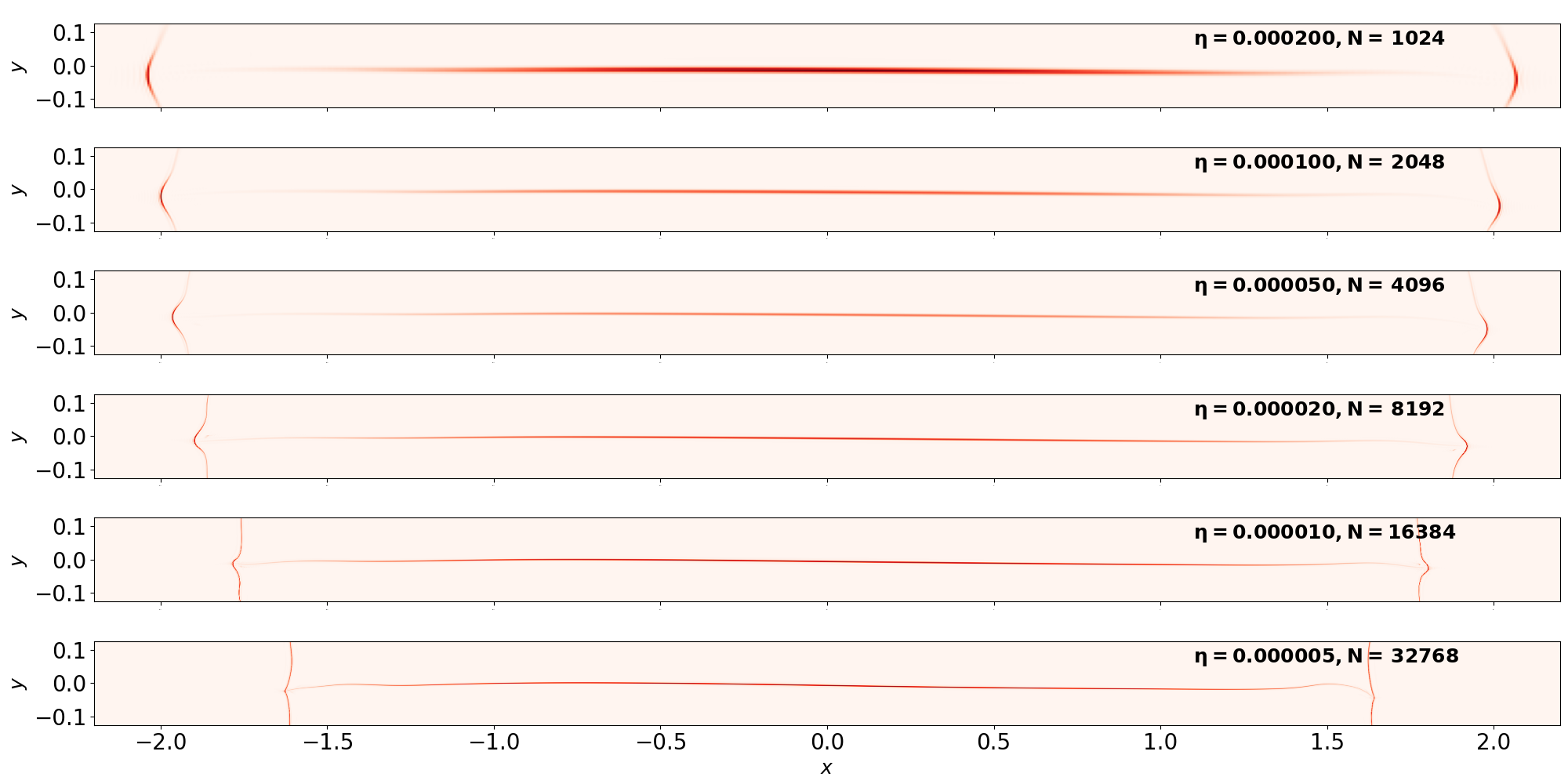} }
  \caption{Squared current density for well resolved runs (zoomed in the current layer) for different values of $\eta$ taken from the marginally well resolved runs. The visualised do main corresponds to the blue box shown in figure \ref{fig:IC}. 
  }
\label{fig:WR}
\end{figure}

\begin{figure}                                                                                   
  \centerline{\includegraphics[width=0.9\textwidth]{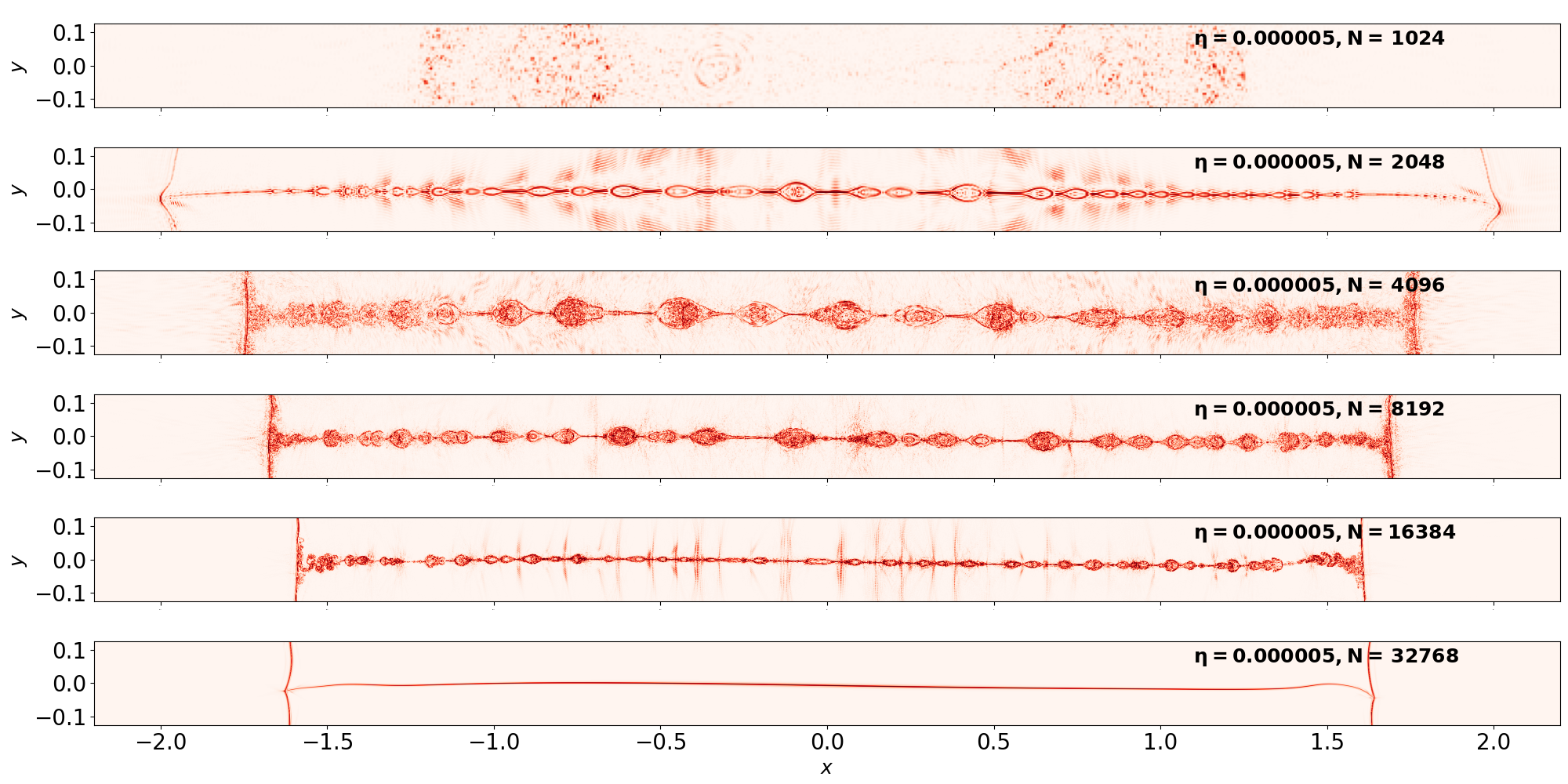} }
  \caption{Squared current density for the smallest value of $\eta$ examined (zoomed in the current layer ) for different resolutions $N$.
  }
\label{fig:UR}
\end{figure}

\begin{figure}                                                                                   
  \centerline{\includegraphics[width=0.48\textwidth]{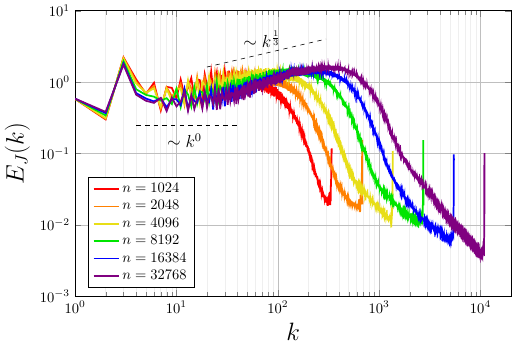} 
              \includegraphics[width=0.48\textwidth]{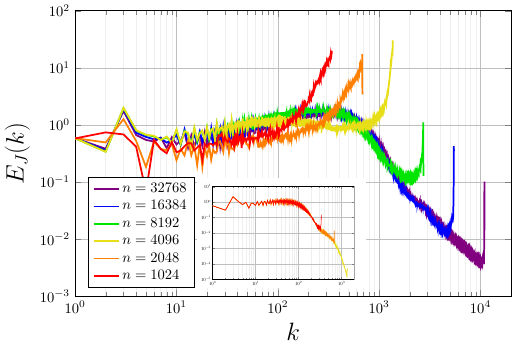} }
  \caption{ Squared current density spectra corresponding to the runs shown in figures \ref{fig:WR} (left) and \ref{fig:UR} (right).
  }
\label{fig:spec}
\end{figure}

%
\begin{figure}                                                                                   
  \centerline{\includegraphics[width=0.48\textwidth]{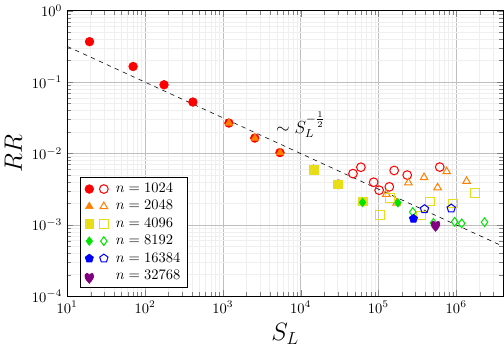} 
            \includegraphics[width=0.48\textwidth]{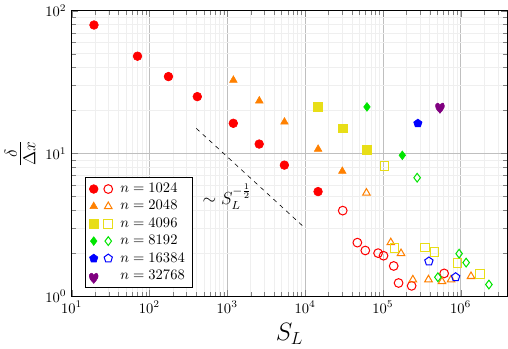} }
  \caption{ Left panel: Reconnection rate as a function of $S_L$ for all runs well-resolved (filled symbols)  and under-resolved (open symbols).
            Right panel: Width of the reconnection layer normalised by the grid size as function of $S_L$.
  }
\label{fig:SP}
\end{figure}

In figure \ref{fig:WR} we show visualisations of the squared current density (zoomed in the current layer)       
obtained from well resolved runs for different values of $\eta$. 
None of these runs displayed visible plasmoids even though values of $S_L=5.4\cdot 10^5$ are reached.
We note that that the current layer is not straight. Instabilities have developed that have 
given a bent shape of the current layer but have not led to plasmoid formation.

In figure \ref{fig:UR} we plot the squared current density again for the smallest value of $\eta$ examined 
for different resolutions $N$. From these runs only the last one for $N=32768$ 
is well resolved based on the criterion mentioned before.
It is striking that all under-resolved runs displayed plasmoids. In fact the worst the
under-resolving the largest the plasmoids appear. This phenomenon is also
present at smaller values of $\eta$ examined: when the well-resolved criterion is violated
plasmoids are present. Table \ref{tab:table2} shows the parameters used for all runs (not just the ones shown in 
figures \ref{fig:WR} and \ref{fig:UR}) indicating the value of resolution $N$ required for each value of $\eta$ 
so that the simulation is well resolved. All resolutions smaller than the marked value displayed plasmoids.
Similar features due to under-resolving have also been observed in Burger's turbulence and the Navier-stokes where they have been studied extensively \citep{ray2011resonance,murugan2023genesis}.  

Further insight can be gained by looking at the energy spectra. The left panel of figure \ref{fig:spec} shows       
the the current density spectra for the runs corresponding to figure \ref{fig:WR}, while the right panel of the 
same figure, shows the spectra for the runs corresponding to figure \ref{fig:UR}. In the left panel all runs are marginally
well resolved. 
As resolution is increased and $\eta$ is decreased $E_J(k)$ progresses to larger wavenumbers forming a $k^0$ power-law range
that reflects the approximate discontinuity of the magnetic field in the current sheet. This power-law range is 
followed by an increase that could be attributed to either
bottleneck 
\citep{falkovich1994bottleneck,donzis2010bottleneck,agrawal2020turbulent}
or a transition to two dimensional turbulence as a result of the instabilities that have developed
with $E(k)\propto k^{-5/3}$. At larger wavenumbers the spectrum shows a steep exponential decrease. 
Finally at the highest wavenumbers near $k_{\max}$ there is a sharp increase.
This is a numerical artifact due to the sharp spectral truncation that leads to a partial thermalisation of the high wavenumbers \citep{cichowlas2005effective,alexakis2020energy}. 
Further increasing the resolution for a given value of $\eta$ has little effect as is demonstrated in the inset
of the right panel where the well-resolved run for $N=1024$ (plotted in the left panel) is repeated  at larger resolutions
$N=2048$ and $N=4096$.                                                                                               

The behavior described above changes when the resolution criterion is not satisfied.                                 
In the right panel of fig. \ref{fig:spec},  where only the simulation with the largest $N$ is well-resolved, 
clear under-resolving features can be testified.
As the resolution is decreased the amount of energy at the largest wavenumbers increases changing the shape of the spectrum.
It is worth noting that the integral of $E_J(k)$ is proportional to the Ohmic dissipation and even at the second to largest 
resolution $N=16\,384$ the Ohmic dissipation due to the wavenumbers at $k_{\max}$ is comparable to the dissipation due to the peak of $E_J(k)$ around $k=500$. It is thus not surprising that violating the well resolved criterion mention before 
can lead to erroneous estimates of the reconnection rate and appearance of plasmoids. 

This is clearly demonstrated in figure \ref{fig:SP} where the reconnection rate $RR$ is plotted as a function of $S_L$
for all our simulations. Filled symbols correspond to well resolved runs while open symbols correspond to 
under-resolved runs. All well resolved runs display the Sweet-Parker scaling $RR\propto S_L^{-1/2}$ even up to 
$S_L= 5 \cdot 10^{5}$ that corresponds to the run at $N=32768$. When the runs are under-resolved however deviations
from this scaling appear, leading to a $S_L-$independent scaling. 
This however is a numerical artifact.
The reason can be linked to the thickness of the
current sheet. In the right panel we plot the thickness of the current layer defined as
$\delta \equiv B_{\max}/j_{max}$ normalised by the grid size $\Delta x=2\pi/N$ for all runs as well.
Well resolved runs follow again the Sweet-Parker prediction $\delta \propto S_L^{-1/2}$ but this scaling
obviously ceases to be true when the width of the current sheet is comparable to the grid size.                        

\section{Conclusions}

Reconnection is a topological change of field lines that can only be broken by micro-scale processes.
Under-resolving can be one of these processes although not a physical one. It is hard to imagine
that continuity of field lines can be preserved when the finiteness of the grid size is apparent. 
Thus care needs to be taken when topological changes are studied with numerical codes. The present 
results indicate that some of the conclusions for magnetic reconnection due to plasmoids in 2D MHD need to be re-evaluated.
We note however that other mechanisms that can lead to a change of field line topology as the ones mentioned 
in the introduction can provide seed to lead to the formation of plasmoids.

\begin{acknowledgments}
This work was granted access to the HPC resources of GENCI-TGCC \& GENCI-CINES (Project No. A0130506421, A0150506421 ). This work has also been supported by the Agence Nationale de la Recherche (ANR project DYSTURB No. ANR-17-CE30-0004 and LASCATURB No. ANR-23-CE30).
\end{acknowledgments}
\bibliographystyle{jpp}
\bibliography{Reconnection}

\end{document}